# An Empirical Study on Decision making for Quality Requirements

Thomas Olsson[1,*], Krzysztof Wnuk[2], Tony Gorschek[2]

[1]RISE SICS AB, Lund, Sweden
[2]Blekinge Institute of Technology, Karlskrona, Sweden

**Abstract.**
[*Context*] Quality requirements are important for product success yet often handled poorly. The problems with scope decision lead to delayed handling and an unbalanced scope.
[Objective] This study characterizes the scope decision process to understand influencing factors and properties affecting the scope decision of quality requirements.
[*Method*] We studied one company's scope decision process over a period of five years. We analyzed the decisions artifacts and interviewed experienced engineers involved in the scope decision process.
[*Results*] Features addressing quality aspects explicitly are a minor part (4.41%) of all features handled. The phase of the product line seems to influence the prevalence and acceptance rate of quality features. Lastly, relying on external stakeholders and upfront analysis seems to lead to long lead-times and an insufficient quality requirements scope.
[*Conclusions*] There is a need to make quality mode explicit in the scope decision process. We propose a scope decision process at a strategic level and a tactical level. The former to address long-term planning and the latter to cater for a speedy process. Furthermore, we believe it is key to balance the stakeholder input with feedback from usage and market in a more direct way than through a long plan-driven process.

## 1 Introduction

Timely deciding on quality requirements (QRs) (also known as non-functional requirements) and accurately setting their expected quality levels (Regnell, Svensson, &

* Corresponding author: Tel. +46 72 550 48 41
*Email addresses*: thomas.olsson@ri.se (T. Olsson), krzysztof.wnuk@bth.se (K. Wnuk), tony.gorschek@bth.se (T. Gorschek)

Olsson, 2008) is challenging but necessary for developing successful software-intensive products (H. B. Kittlaus & Clough, 2009), especially products released to open consumer markets (Regnell, Svensson, & Olsson, 2008). Our previous work indicates that quality aspects play an important - even dominant - role in a product purchase decision for MDRE (Regnell, Svensson, & Wnuk, 2008). Despite that, QRs are often poorly understood, informally stated and difficult to validate (Berntsson Svensson et al., 2012; Chung, Nixon, Yu, & Mylopoulos, 1999). Moreover, QRs are often *subjective* (can be evaluated and interpreted differently), *relative* (some level of quality is always reached) and *interacting* (one quality aspect interacts and influences positively or negatively another aspect) (Berntsson Svensson, Olsson, & Regnell, 2013; Chung et al., 1999).

This paper focuses on scope decisions (Wnuk & Kollu, 2016) of features across the lifecycle of a products line (H. B. Kittlaus & Clough, 2009). Rather than making scope decisions on individual requirements, decision makers typically group requirements into features (Gorschek & Wohlin, 2006). We analyze five years of scope decisions history (from 2010 to 2015) with the objective of identifying quality features (QFs) in the scope decision process and exploring how scope decisions are made for those. We study how decisions on QFs are made in the scope decision process; how different stakeholder organizations are involved; when decisions are made in relation to the release plan, etc.

The goal is to characterize the scope decision process for QFs to understand influencing factors and intrinsic properties. This paper addresses the research gap by explicitly focusing on scope decisions across several releases and products. We analyzed the scope decision history between 2010 and 2015 of 4446 features at the case company, including more than 41 products and 36 software releases of a product line. We identified 196 QFs and performed interviews with senior practitioners at the company confirm our interpretations from the observations from the scope decision analysis and to understand the rationale behind the decisions.

This paper is organized as follows. The background and related work are discussed in Section 2. A description of the case study company is found in Section 3. Section 4 outlines our research methodology and research questions. Our analysis and results are presented in Section 5. Section 6 contains discussions and implications of the results. Finally, the paper is concluded in Section 7.

## 2    Background and Related Work

### 2.1    Background

One of the main challenges with QRs during the Requirements Engineering (RE) process is to involve relevant stakeholders holding various roles and achieve an agreement on *what* to do and *when* to do it (Franch, 1998). This is usually referred to as requirements scope decision (Wnuk & Kollu, 2016). (Sometimes simply referred to as scoping.) The requirements scope decision process is a continuous decision activity performed on multiple-levels aimed at finding a scope that satisfies as many needs and constraints as possible. Scope decisions support software product managers in translating the product strategies into a series of software releases via deciding what features to realize in a software release or a software product (Wnuk & Kollu, 2016). Scope decisions can be divided into strategic (deals with goals and objectives), tactical (identification and use of sources and resources) and operational (assuring effectiveness when performing the operations) (Anthony, 1965; Aurum & Wohlin, 2003).

Market-Driven Requirements Engineering (MDRE) increases the importance of releasing the right product to the right market at the right time (Regnell & Brinkkemper, 2005) and strengthens the importance of decision making of QRs. The continuous nature of requirements scope decision plays an important role in bridging strategic product portfolio planning and associated release planning with operational scope decisions that need to be taken to adapt to unexpected changes (Wnuk et al., 2016). The software product management literature (H. B. Kittlaus & Clough, 2009) recognizes the strategic importance of QRs in setting the product strategy (ISPMA, 2014) but does not consider its special nature during the product planning and release planning processes. Moreover, software roadmaps that provide the input for release planning mostly focus on planning how to use available technological resources and scientific knowledge over a series of product releases (Regnell & Brinkkemper, 2005; Vähäniitty, Lassenius, & Rautiainen, 2002). Roadmaps often focus on technology (Kostoff & Schaller, 2001; Rinne, 2004) or product-technology aspects rather than quality and user experience aspects. Others highlight that to achieve customer value it is more important to connect the strategic roadmap to customer value rather than on the prioritization of individual features (Komssi, Kauppinen, Töhönen, Lehtola, & Davis, 2015).

We define features specifically aimed at improving a specific quality aspect (QA) as quality features (QFs) in this paper. The QFs are contrasted by functional features (FF), where the focus is primarily on adding new functionality and quality aspects remain embedded. A product family is a grouping of products for marketing reasons (ISPMA, 2014; H. B. Kittlaus & Clough, 2009). A product line is a systematic way to reuse features across different products and systems (H. B. Kittlaus & Clough, 2009). A platform is a set of common software component and surrounding infrastructure on which products are developed (Pohl Klaus, Böckle Günter, & Van der Linden Frank, 2005). We use the term product line variant to refer to the parallel development of the product line which takes places on different projects. Lastly, a (software) product is based on a product line variant and is the software which is intended for a specific hardware product.

## 2.2 Related work

One of the earliest empirical papers in the RE literature with results related to QRs is the study by Lubars et al., (Lubars, Potts, & Richter, 1993). The main challenges identified in this work are vague requirements, changing requirements and difficulties to prioritize. A later survey on QRs highlights interdependencies and scope decision of QRs as significant challenges (Berntsson Svensson et al., 2012). Several researchers highlight the importance of QRs during software product development. Disregarding

QRs might lead to a product that is too difficult to use or too expensive to maintain (Ebert, 1998). Moreover, poor management of QRs might lead to project overruns and increased time-to-market (Cysneiros & Leite, 2004). A report from the automotive industry indicates that engineers agree that QRs are far more important than functional requirements (FRs) but the competence to specify them is lacking (Weber & Weisbrod, 2002). In our previous work, we have seen that QRs are particularly important to market-driven organizations that release products to a consumer market (Regnell, Svensson, & Olsson, 2008). Although the importance of QRs is generally acknowledged, Cysneiros and Leite report that RE research is dominated by functional requirements (Cysneiros & Leite, 2004), whereas Ameller et al., (Ameller, Ayala, Cabot, & Franch, 2013) report that both FR and QRs are considered equally important by software architects. Hence, there is not an agreement in the community. In this study, we complement the current body of knowledge with an in-depth view of release planning of QRs in an MDRE context where we specifically study the input to the scope decisions for QRs.

A recent literature review on QRs for SPLs suggest that there is a lack of understanding for trade-offs among QRs and that there is a need to improve the alignment between QR and SPL practices (Soares, Potena, Machado, Crnkovic, & De Almeida, 2014). Others have studies performance variability and found that the rationale for varying the performance is both price (customer-driven) and actual usage (solution-driven) (Myllärniemi, Savolainen, Raatikainen, & Männistö, 2016). Furthermore, they stress the need to improve performance requirements over time, as the systems in their study are very long-lived and that different QRs require different solutions regarding variability in an SPL. We analyze the scope decisions for one SPL

With the data-driven software engineering trend (Bosch, 2016), it is suggested that it is better to perform measurements and incrementally change the scope through continuous experimentation (Parnin et al., 2017). However, as Parnin et al., point out, not every feature warrants an experiment. Studies suggest that to work data-driven requires, e.g., a suitable infrastructure and a connection between products roadmaps and experiments (Fagerholm, Sanchez Guinea, Mäenpää, & Münch, 2017). Furthermore, organizational factors and complex stakeholder structures inhibit continuous experimentation (Yaman et al., 2017). In our study, we analyze the scope decision process in a context where data-driven ideas are introduced over the years and how scope decisions for QRs are related to data.

Incremental delivery of software gains importance and impacts release planning methods and processes (Ruhe & Saliu, 2005). Interesting work has been done to combine the experience of individuals with computer-supported tools in release planning (Greer & Ruhe, 2004). Release planning for QRs specifically was studied by Regnell et al., (Regnell, Svensson, & Olsson, 2008) while Carlshamre et al., focus on interdependencies among requirements in software release planning (Carlshamre, Sandahl, Lindvall, Regnell, & Natt och Dag, 2001). Berntsson Svensson re-used the interdependency types suggested by Carlshamre et al., to study dependencies between QRs but without a clear release planning angle. Heikkilä et al. studied release planning in a mixed agile and plan-driven context (Heikkilä, Paasivaara, Lasssenius, Damian, & Engblom, 2017). Conclusions are that the combination if different process philosophies can reduce lead-time, improve flexibility and planning efficiency but identify difficulties in for example organizing system-level work and balancing effort between different processes. Our work focuses on how to plan and deliver QRs across many releases, with each release taking the software closer to the fulfillment of the complete QR and the need to combine both long-term, plan-driven processes with fast and agile processes, confirming that other results are also relevant for QFs. However, our findings point to the importance of considering analytics and BI to be able to timely implement QRs and to set the quality level appropriately.

A survey on how software architects see QRs concludes that the QRs are important for them and there is a lack of tools and customizable methods which can be used in practice (Ameller et al., 2013). Ernst and Mylopoulos report that the handling of QRs in the lifecycle varies and they found no correlation with the age of the project in their study (Ernst & Mylopoulos, 2010). Also, Cleland-Huang et al., notice that QRs typically are discovered late in software development projects, and often in an ad hoc fashion (Cleland-huang, Settimi, Zou, & Solc, 2007). Borg et al., identified a number of QR challenges in a case study (Borg, Yong, Carlshamre, & Sandahl, 2003). The main finding was again that QRs are discovered late - if they were discovered at all. We complement the existing important empirical results by providing a longitudinal case study, analyzing the handling of QR over a product line lifecycle of five years.

## 3     Case company

The case company develops software-intensive products for a B2C global market. The company is a large multi-national company with a long history, stretching back to the first half of the 20$^{th}$ century. For the specific product line, the number of engineers varied from 1000 to 4000 worldwide (developers, project managers, product managers, testers, etc.) during the studied period.

We selected this case as the company has a long history of working with scope decisions and has experienced challenges with QRs. Furthermore, we had a unique opportunity to access and analyses the scope decision database.

### 3.1     Object of study

The studied product line constitutes hand-held and battery powered products. The products have extensive networking capability, high computational power and can be extended with the third-party software applications. The products are sold on a global market to both private consumers as well as to companies.

The company develops 4-8 unique products each year on this product line, where each product has a volume of around 5-20 million units. Each product has around 2-5 major software releases and additionally 5-10 minor to address variants and versions for specific market needs. The major releases include the new features. The minor releases are primarily maintenance releases to fix defects and to release the software to additional markets. In this study, we focus on the major releases, as our interest is in the scope decision process for new QFs.

The product development projects typically run over 12-18 months and include around 400-500 software engineers. In parallel, there is a team of 50-100 software engineers responsible for the product software unique for a specific product, e.g. unique hardware. For the common software across the product line, there is a separate organization. From early on, a software product line engineering approach was established (Pohl Klaus et al., 2005). In **Fig. 1**, a product is illustrated with the large arrows, one for each product. The products are called H1, M2[1], etc. in the figure. The H product family is the high-end products and the M family mid-end products in the example. The numbers are identifiers. There are also low-end products in the totally different product family, based on an entirely different product line, and therefore not included in the study. A product line variant is illustrated with the group of arrows in the same pattern. There is one project per product line variant.

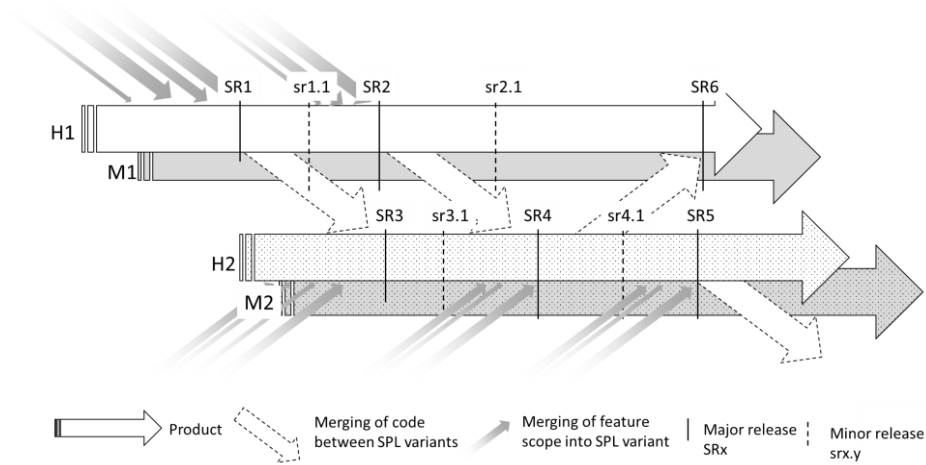

**Fig. 1.** An overview of the Software Releases (SR) and software dependencies across product line variants. High-end products are denoted H with a number indicating order (e.g., H1) and mid-end products M in a similar fashion (e.g., M2).

The projects handle a combination of bespoke and market-driven requirements. The products are developed for a B2C mass-market, i.e. MDRE (Regnell, Svensson, & Wnuk, 2008). Most of the feature suggestions come from internal stakeholders such as technical experts or marketing managers. However, there are partners which have extensive influence on the scope decisions. These external stakeholders interface the scope decision process through a customer account organization. In total, it is a very complex and large scope ranging from very technical parts to user experience related features. Individual software features, such as adding a new networking capability, are developed in smaller sub-projects consisting of around 3-15 software developers and a software architect. A feature project typically is executed over a period of 2-8 weeks. When they are finished, the individual features are included in the product line.

---

[1] For confidentiality reasons, the product identifiers are anonymized.

The release planning process is one major decision process for the Software Product Manager (SPM) (H. B. Kittlaus & Clough, 2009; Ruhe & Saliu, 2005). For the case company, there is one product that leads a major release of the product line, where the main development for the common software is performed, see **Fig. 1**. A major release (e.g., SR1 in **Fig. 1**) can, for example, include an upgrade of the underlying operating system, a large amount of new functionality or large overhauls of the UI. The main software development investment is spent on the major releases. Besides the major releases, there are minor releases (e.g., sr1.1 and sr1.2 see **Fig. 1**) with fewer changes and fewer investments. Typically, one part of the product line software project is dedicated for a release in a release (sub-)project. The release projects run in parallel and often overlap regarding software development effort and scope decision, both with a product line variant and across the entire product line. Hence, the release planning context is complex with many parallel products and projects and many decision-makers who need to align.

### 3.2 Major events during the studied period

The scope of the study is software scope decisions for the product line during the period from 2010 to 2015. In 2010, when the product line was initiated, the market was immature and expanding, see **Fig. 2**. In this build-up phase, much of the strategic focus was technology driven, to introduce new hardware and software features to the market. After the build-up phase, the product line and markets grew substantially in the growth phase. At the beginning of the product line lifecycle, approximately 1000 developers were working on the product line. At the time, this constituted about 25% of the software engineering resources for the entire product portfolio. At the beginning of the product line lifecycle, there was no systematic approach to the product line scope decisions nor reuse. However, early on, product line concepts were introduced to be able to cope with the growth of the product family and the expanding market.

| Strategy focus | Technology | | | | User experience |
|---|---|---|---|---|---|
| Market characteristics | Immature | Expansion | | Consolidation | |
| Product life-cycle phase | Build-up | Growth | | New Markets | Consolidate |
| Development paradigm | Product-focus No SPL | Platform-focus SPL | | Product-focus SPL | |
| Engineers | 1000 (25%) | 4000 (100%) | | 2500 (100%) | |
| | 2010 | 2011 | 2012 | 2013 | 2014 | 2015 |

**Fig. 2.** Main characteristics of the product line during 2010-2015

The market matured around 2013 shifting the focus on user experience and differentiation rather than functionalities. This signifies the change to the new market phase.

This is apparent for the last 2 years of the studied period. The number of developers increased to approximately 4000 engineers 2/3 into the 5-year period, which at the time constituted almost 100% of the software development resources. At the end of the studied period, the number of engineers involved in the development process was approximately 2500. The final year of the studied period is signified by a consolidation phase, where the focus is on development efficiency and optimization of reuse utilization.

The growth focus of the product line with a more advanced platform to allow for the many products and continued innovation. From late 2013, there was a shift in focus from supporting a large and expanding product line to development resource efficiency across the product line, i.e., being able to maintain and improve the product line with less personnel. In parallel, the scope decision process is changed to a stronger product focus from previously a platform focus.

### 3.3 The software scope decision process

We study the scope decisions for software features across the product line lifecycle for features. A feature, in this case, is a scope decision entity that contains between 20 and 50 detailed requirements (Wnuk et al., 2016) and is intended to address a specific need and add a business value if realized (Gorschek & Wohlin, 2006; Regnell & Brinkkemper, 2005). An FF can be to add or extend existing functionality, e.g., adding support for new audio formats. A QF is a feature explicitly targeting improving a specific QA, e.g., battery consumption while playing audio.

An SPM is responsible for the software for a specific product. For example, in **Fig. 1**, there is one SPM for H1, one SPM for M1, etc. Hence, for one software project, there are several SPMs. There is typically one product which leads a variant of the product line and is responsible for aligning the decisions among all the products using that product line variant. The SPM follows the product from its inception to the end-of-life.

The major decisions for an SPM are on features, both whether to include or not as well as when to release these features Decisions are taken continuously, i.e. there is no specific phase or milestone when features are submitted to the SPM or decisions are made.

## 4 Research overview

The goal of our study is to characterize the scope decision process. The scope decision process is a complex and multi-faceted process. A flexible research design is suitable for studying these kinds of phenomena (Robinson, 1993). Importantly, we collected both quantitative and qualitative data to get a holistic view. Specifically, we decided to utilize an exploratory case study research strategy (Runeson, Höst, Rainer, & Regnell, 2012) which is applicable to flexible research designs. Runeson and Höst state that case studies are well suited for many types of software engineering research. Seaman (Seaman, 1999) stressed the ability of case studies to grasp the complexity of the studied problem while Wieringa and Heerkens (Wieringa & Heerkens, 2007) considered case studies as suitable for requirements engineering research. By studying one company in depth, we can discover phenomena and explore patterns of scope decision and release planning of QFs.

## 4.1 Research questions

As outlined in Section 1, our goal is to characterize the scope decision process. **Table 1** outlines the research questions investigated in this paper.

**Table 1**
Research questions

| Research question |
| --- |
| **RQ1.** Are quality aspects explicitly specified for features? |
| **RQ2.** What does the scope decision and scope decision process for quality features (QFs) look like across the product line lifecycle? |
| **RQ3.** How do different roles influence the decision for quality features (QFs) over the scope decision process lifecycle? |

The first research question (**RQ1**) aims at understanding whether there are QFs present in the scope decision process. Related work by Eckhardt et al. (Eckhardt, Vogelsang, & Fernández, 2016) suggest that there is a reason to treat QRs in the same way as functional requirements. Furthermore, in our work previous work (Berntsson Svensson et al., 2013), we have investigated the prevalence of QRs in a requirements specification. With **RQ1**, we investigate QFs in the scope decision process, which to our knowledge has not previously been studied.

Research question **RQ2** is inspired by the work of Berntsson Svensson et al., (Berntsson Svensson et al., 2012) who qualitatively surveyed several companies asking about managing QRs. They focused on which quality aspects are the most important (also explored in our work (Berntsson Svensson et al., 2013)), how to handle interdependencies between them and how to estimate their cost. Furthermore, they discovered that 19% of all QRs are dismissed at some stage before release. In previous work, we have found that change requests from external stakeholders are more likely to be accepted (Wnuk, Kabbedijk, Brinkkemper, Regnell, & Callele, 2015). We have previously concluded that features proposed late in the release cycle are more often rejected than accepted (Wnuk et al., 2015). However, they have not focused primarily on QRs or QFs and did not study what happens across several releases in the product planning process. Ernst and Mylopoulos (Ernst & Mylopoulos, 2010) mined eight open source projects exploring if QRs are mentioned more often as the projects progress. They found no correlation with age or lifecycle regarding the importance of QRs. The novelty of our study lays in the fact that it presents the detailed view of how QFs are handled in product line lifecycle and in the different stages of the scope decision process. Our study is based on a manual analysis of historical data on the scope decision process and not dependent on data mining techniques.

Research question **RQ3** focuses on the different stages in the scope decision process, with an emphasis on the decisions made by the SPM role and influence from stakeholder roles – whether internal or external. Ebert stresses the importance of the SPM role to ensure a lifecycle perspective and bridge perspectives across products and stakeholders (Ebert, 2007). Furthermore, Ebert and Brinkkemper stressed the critical role of portfolio management for road-mapping of software products to facilitate transparency and dependency management (Ebert & Brinkkemper, 2014). Knauss et al., (Knauss, Damian, Knauss, & Borici, 2014) discovered the importance of strategic flow and involving the relevant stakeholders in flow management at IBM; whether a specific feature is included in the scope is typically up to an SPM. We complement the existing work with a detailed analysis of the decisions on QFs for the different decision makers in the scope decision process. We also analyze how internal stakeholders – representing experts within the company – influence the decisions compared to external stakeholders – representing organizations outside the company – and how it varies over the portfolio lifecycle.

## 4.2 Research method

We performed an archival analysis of documentation data from an electronic database with extensive meta-data (Lethbridge, Sim, & Singer, 2005). We extracted data from the scope decision database that contained information on features and release planning. The database is rich and extensive which enable an in-depth analysis not dependent on peoples recollection of the events (Robinson, 1993). However, the downside is that the information was not stored to support this research endeavor. Hence, the data requires a fair amount of pre-processing before it can be analyzed. Furthermore, it is not possible to understand the rationale and underlying reasoning. For this purpose, we also performed interviews with experienced engineers involved in various aspects of the scope decision process for many years. The interviewees complement the archival analysis with explanations and rationales as well as missing data.

The aim of the interviews was to complement the quantitative data with qualitative interviews with persons who have worked with the scope decisions. The interviews were semi-structured. The structured part was formulated to confirm specific interpretations of the data to ensure we had understood the data correctly. The unstructured, open-ended part was a more open dialogue to capture additional findings not necessarily visible in the decision database.

The data were analyzed using descriptive and inference statistical methods to explore correlations and to test assumptions. As the underlying distribution of the various data is unknown, we used non-parametric statistics (Sheskin, 2004). We used the Wilcoxon signed-ranks test to test assumptions and boxplots to visualize variance in the data along with common bar plots. The anonymized raw data used for statistical testing was packaged and can be obtained upon request.

## 4.3 Research process and data collection techniques

Our research process was divided into two research phases (cf. **Fig 3**):

1. **Quantitative archival analysis** of the scope decision database
2. **Qualitative interviews** with senior engineers at the case company.

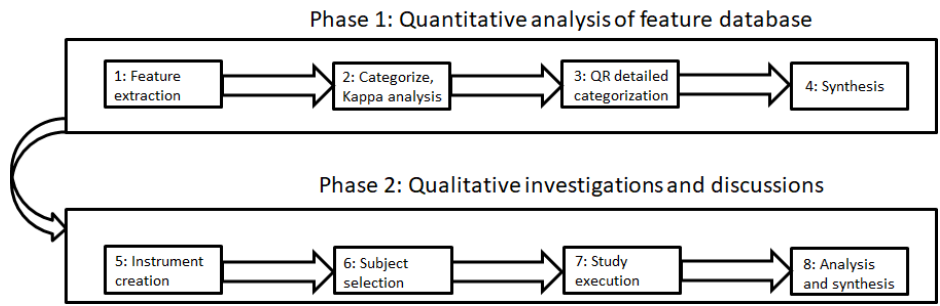

**Fig 3.** Research process overview.

**Research phase 1:** The core of the first research phase is a quantitative analysis of scope decision archive at the case company. We conducted this is 4 steps:
1. We extracted all the features from the feature decision database for the product line in question, see Section 3.1. We removed duplicates and administrative features. **Table 2** outlines the extracted type of data.
2. We categorized the extracted relevant features as either QFs (according to ISO25010 (International Organization For Standardization (ISO), 2011)) or FFs. Two researchers were involved in the categorization to ensure consistency and discuss discrepancies. First, the first author categorized all features and the second author sampled 10%. Analyzing the disagreements by means of Cohen's Kappa for interrater agreement (Cohen, 1968) resulted in a value of 0.75. After discussing and agreeing on the categorization, we reached an interrater agreement above 0.90. However, as only 10% of the features were sampled and only 4.41% of the QFs, we concluded that all features should be coded by both the first and second author to ensure reliable categorization. After the first round of categorization, 194 features were categorized as QF; after the second round 196 (2 QFs from round 1 were FFs, and 4 FFs from round 1 was in fact QFs.)
3. After having identified all the QFs, we extracted additional meta-data such as lead-time in various states and manual analysis of minutes of meetings, etc. The manual coding was performed by one researcher and the other reviewed it. This was iterated until 100% alignment was achieved. **Table 3** describes the additional meta-data that was collected. 650 features (both FFs and QFs) had extreme lead-time, likely a result of unreliable data. We identified the unreliable lead-times by manually examining features with very long lead-time. If the lead-time for a feature is more than 365 days of total lead-time, they were subject to removal. We sampled 10% of the features to conclude whether the cut-off value of 365 days was appropriate. In all but 2 of the sampled features, the cut-off value made sense. We, therefore, concluded that removing the lead-time for all features where one or more of the parts exceeded 365 days was sensible.
4. The data was analyzed to find trends and synthesize conclusions.

**Table 2**
Data extracted from the feature decision database

| Type | Description |
| --- | --- |
| **Title** | A sentence, typically 2-10 words |
| **Description** | Typically, 1-3 paragraphs of text |
| **Origin** | Which stakeholder proposed the feature (customer, internal stakeholders, etc.) |
| **Priority** | Priority as assigned by the SPM |
| **Minutes of Meeting** | A log of all decisions taken in the scope decision process |
| **Type of Feature** | New feature or administrative entries in the database |

**Table 3**
Meta-data collected

| Type | Description |
| --- | --- |
| **Type of QF** | According to ISO25010 |
| **Proposed release** | As suggested by the stakeholder who proposed a feature |
| **Planned release** | As decided by the SPM |
| **Actual release** | The software release where the QF is implemented |
| **Lead-time in various states** | How long a feature is in a state as prescribed by the scope decision process at the case company. |

**Research phase 2:** The second research phase included interviews with senior engineers at the company directly involved in the scope decision processes. The aim was to complement the quantitative data with qualitative interviews with persons who have worked with the scope decisions.
5. We created an interview instrument during several discussion sessions between the researchers. The interviews were semi-structured and focused on providing qualitative information to confirm observations from the archival analysis and to further explain the underlying behavior. The interview instrument is available in Appendix A.
6. The subjects were selected based on their roles in the scope decision process for the product line in question. They were also selected to have varying perspectives on the scope decision process to cover as possible.
7. The interviews were performed by two researchers. The interviews took on average one hour. The questions in the interview instrument were discussed and documented in interview notes. The final interview questions are found in Appendix A.

8. Interviews were transcribed and analyzed using content analysis (Robinson, 1993). The transcripts were analyzed by the first two researchers who mapped them with interview questions and marked interesting sections. Next, we examined qualitative data searching for different perspectives on managing QFs at the case company.

## 4.4 Execution

The data was collected in mid-2015 and the subsequent interviews conducted at the beginning of 2016. In total, 4 interviews were performed with persons who have had different roles related to the scope decision process over the years. All the interviewees have over the years been involved in central processes around scope decisions both on an operational and strategic level. In total, there are less than 25 persons at the case company with similar insights and experience as the interviewees. Their roles and perspectives are found in **Table 4**.

**Table 4**
Interviewees' roles, experience, and additional information.

| Interviewee | Role | Years at company | Comment |
|---|---|---|---|
| **I1** | SPM | 8 | I1 worked in the role of an SPM from 2008-2014. I1 was one of the original SPMs involved in the product line analyzed in this case. |
| **I2** | PO | 8 | I2 has worked as a product owner (PO) for different sub-areas within the product line. As PO for an area, I2 was involved as an internal stakeholder to the product line from the beginning of the lifecycle and is still involved. |
| **I3** | Architect | 10 | I3 is a line manager for the systems architecture department. The system architects staff the development projects with senior architects responsible for impact analysis of new features, architectural reviews, contribution strategy, etc. |
| **I4** | Customer account and SPM | 8 | I4 worked as customer account representative for external stakeholders between 2010-2011 and as an SPM for the product line in question during 2012-2015. In addition. I4 was also line manager for the SPM group during 2014. |

## 5 Results and Analysis

We extracted 7116 features from 36 software releases and 60 products between 2010 and 2015. After initial screening, 41 products over 10 different chipsets and 36 software releases concerning 5398 features were included. Next, we removed administrative features used to keep track of development resources across the releases, keeping 4446 features. The features were furthermore classified as either FF or QF. We identified 196 QFs from the 4446 features (4.41%) that have been discussed at some point in the scope decision process. **Table 5** summarizes the features extracted and included in the scope of the study.

**Table 5**
Features extracted and included in the scope of the study.

|  | Amount |
|---|---|
| **Features extracted** | 7116 |
| **After removing out of scope products** | 5398 |
| **After removing administrative features** | 4446 |
| **FF included** | 4250 |
| **QF included** | 196 |

We identified 6 major findings in the data. Firstly, the presence of quality aspects in the scope decision process, presented in section 5.1. In section 5.2, findings related to the lead-time of features are presented and in section 5.3, findings on acceptance ratio in relation to the stage in the scope decision process are presented. An analysis of how the type of stakeholder influence the scope decision process for QAs is presented in section 5.4. Findings related to the release planning process in relation to QAs in the scope decision process are presented in section 5.5. Lastly, findings related to feedback on the decided quality level are presented in section 5.6.

### 5.1 Presence of quality aspects in the scope decision process

Of the 4446 features included in the analysis, 196 are primarily focused on quality, see **Table 5**. We further classified the identified 196 QFs according to ISO 25010 (International Organization For Standardization (ISO), 2011), summarized in **Fig. 4**. As can be seen, if considering all QFs for all releases, 48% are of the performance QA. The second largest QA of QFs is security constituting 30% of the QFs, followed by Usability being 13% of the QFs. Hence, in relation to RQ1, the results clearly show that quality aspects are in fact explicitly specified and part of scope decisions.

| | ISO 25010 category | Number of QFs | % of total QFs |
|---|---|---|---|
| Performance | Capacity | 3 | 2% |
| | Resource utilization | 64 | 33% |
| | Timing behaviour | 27 | 14% |
| | Total | 94 | 48% |
| Security | General | 2 | 1% |
| | Authentication | 1 | 1% |
| | Authenticity | 17 | 9% |
| | Integrity | 39 | 20% |
| | Total | 59 | 30% |
| Usability | Accessibility | 2 | 1% |
| | Appropriateness recognisability | 4 | 2% |
| | Operability | 20 | 10% |
| | Total | 26 | 13% |
| Other | Fault tolerance (Rel.) | 7 | 4% |
| | Maturity (Rel.) | 3 | 2% |
| | Functional appriopriateness (Func.) | 2 | 1% |
| | Analysability (Maint.) | 2 | 1% |
| | Adaptability (Port.) | 2 | 1% |
| | Recoverability (Rel.) | 1 | 1% |
| | Total | 17 | 9% |
| | **Total number of Qfs** | **196** | |

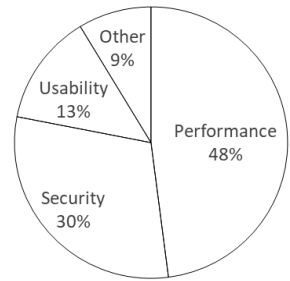

**Fig. 4.** Overview of the classification of the QFs according to ISO 25010 (International Organization For Standardization (ISO), 2011) for the three most prevalent categories (Performance, Security and Usability) and other represented sub-categories from reliability, maintainability, portability and functional suitability.

In the build-up phase, the products were defined with little reuse of software among them since the focus was on introducing products to the market as fast as possible. Hence, a strong focus on application engineering (Pohl Klaus et al., 2005), and an evolutionary approach to the product line initiation (Bosch, 2002). The product line is initiated and build-up based on an old configuration (similar to experiences reported by Deelstra, Sinnema, and Bosch (Deelstra, Sinnema, & Bosch, 2005)), in a bottom-up manner (similar to experiences from Li and Chang (Li & Chang, 2009). This phase is also characterized by learning the new technology and markets. Hence, related to RQ1, the data indicates that even though quality aspects are explicit in the decisions, the prevalence varies across the lifecycle.

**Table 6**
Summary of functional (FF) and quality (QF) features in each lifecycle phase

| Phase | Release | FF | QF | Total | % QF |
|---|---|---|---|---|---|
| Build-up | 2010Q2 | 295 | 10 | 305 | 3.3% |
| | 2010Q3 | 104 | 1 | 105 | 1.0% |
| | 2010Q4 | 92 | 1 | 93 | 1.1% |
| | Total | 491 | 12 | 503 | 2.4% |
| Growth | 2011Q1 | 280 | 5 | 285 | 1.8% |
| | 2011Q2 | 219 | 6 | 225 | 2.7% |
| | 2011Q3 | 168 | 1 | 169 | 0.6% |
| | 2011Q4 | 24 | | 24 | 0.0% |
| | 2012Q1 | 218 | 10 | 228 | 4.4% |
| | 2012Q2 | 50 | 2 | 52 | 3.8% |
| | 2012Q3 | 244 | 4 | 248 | 1.6% |
| | 2012Q4 | 126 | 1 | 127 | 0.8% |
| | Total | 1329 | 29 | 1358 | 2.1% |
| New market | 2013Q1 | 268 | 10 | 278 | 3.6% |
| | 2013Q2 | 61 | 9 | 70 | 12.9% |
| | 2013Q3 | 290 | 18 | 308 | 5.8% |
| | 2013Q4 | 50 | 8 | 58 | 13.8% |
| | 2014Q1 | 4 | | 4 | 0.0% |
| | Total | 673 | 45 | 718 | 6.3% |
| Consolidation | 2014Q2 | 453 | 27 | 480 | 5.6% |
| | 2014Q3 | 332 | 44 | 376 | 11.7% |
| | 2015Q1 | 96 | 1 | 97 | 1.0% |
| | 2015Q2 | 484 | 23 | 507 | 4.5% |
| | 2015Q3 | 367 | 15 | 382 | 3.9% |
| | 2015Q4 | 25 | | 25 | 0.0% |
| | Total | 1757 | 110 | 1867 | 5.9% |
| | Total | 4250 | 196 | 4446 | 4.4% |

**Table 6** depicts how submitted FF and QF in each phase. Both the build-up and the growth phases have few QFs. In the build-up and growth phases, the median relative number of QFs (% QF in **Table 6**) is 1.6%, whereas the median number of QFs is 4.6% QFs in the new market and consolidation phases, i.e. a tripling of the ratio of QF/FF. If we test the relative number of QFs per release in the build-up and growth phase compared to the new market and consolidation phase using a Wilcoxon signed-ranks test, we get a result of $p = 0.057$, i.e. statistically likely that there is a difference between the two samples. The new market phase has a larger ratio of QRs (5.9%) to the total number of features while the build-up phase has the lowest ratio (2.4%). However, the difference in the ratio of QFs across the phases is not significant if using a Wilcoxon signed-ranks test. **Fig. 5** visualizes how the relative number of QFs varies across the phases.

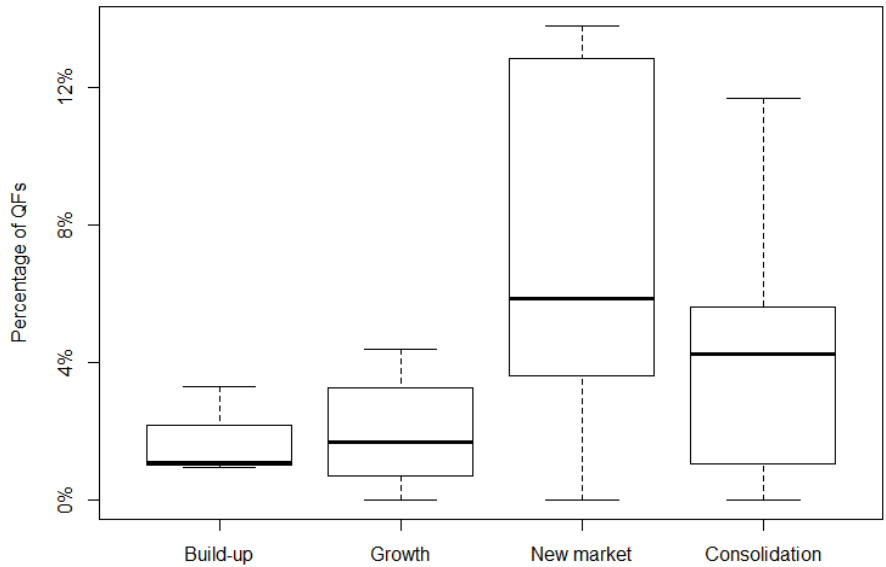

**Fig. 5.** The relative number of QFs (denoted % QF in **Table 6**) for the different releases across the four phases.

In the build-up phase the focus, in terms of QFs, is on performance (58% of the QFs in the build-up phase). Interviewees I1 and I3 pointed out that they perceived a focus on FF, both from strategic planning as well as scope decision perspective, albeit not formalized in any document. Interviewee I1 commented that battery performance has always been a problem, even in the build-up phase. Interviewee I2 mentioned that for the areas they are responsible for they did discuss this but never perceived any overarching strategy.

There was no variability handling of QAs in the product line. A possible interpretation of our result is when starting a product line utilizing agile software development, quality is neglected, and functionality is prioritized.

### 5.2 Lead-time of features from inception to end-state

In the growth phase, the software started to be developed with a software product line approach (Pohl Klaus et al., 2005). The existing separate software tracks were merged, resulting in significant architecture refactoring work for creating high-quality and stable domain artifacts. This was a preparation phase for the planned growth of the product family in the number of products and markets. This significantly affects the software development as well, where the focus during the growth phase is on the domain artifacts and not the application artifacts (Pohl Klaus et al., 2005). The challenges faced were similar to that of the Boeing case (Pohl Klaus et al., 2005), i.e. harmonizing different components and architectures into a domain architecture. The increased complexity is visible through an increased lead-time for making a scope decision for a feature (both FFs and QFs) from 7 days in the median for the build-up phase to 59 days for the growth phase (significant at p = 0.1 using a Wilcoxon signed-ranks test). The increase for QFs is even greater, where decision lead-time increases from 7 days to 65 days. Note that these numbers are for all features, independent of end-state. If, for example, a feature is discussed but never defined the lead-time is sometimes short.

**Fig. 6** presents boxplots for the lead-time for the features across the phases. The lead-time to discard a feature is, in general, longer than to complete the implementation of a feature. We do in this case not distinguish when a feature is discarded – just the fact that it is discarded rather than completed (cf. **Fig. 7**). The median lead-times for completed features are 91%, 109%, 21% and 44% of the median lead-times of discarded features. The difference is significant when testing it using a Wilcoxon signed-ranks test at p=0.1. There is no significant difference in lead-time when considering the difference between FFs and QFs (cf. RQ2).

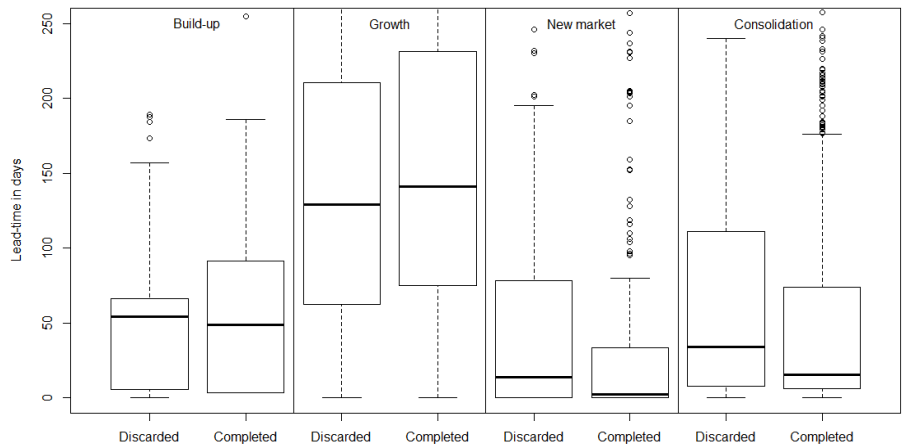

**Fig. 6.** Lead-time in days across the phases and whether completed or discarded

### 5.3 Acceptance ratio across the stages

The scope decision process of the case company consists of three major stages, as outlined in **Fig. 7**. A) The IN stage where stakeholders are formally discussing a proposal,

B) the PLAN stage, where the product management is considering a feature, and C) PROJECT stage where there is concrete design and implementation work performed.

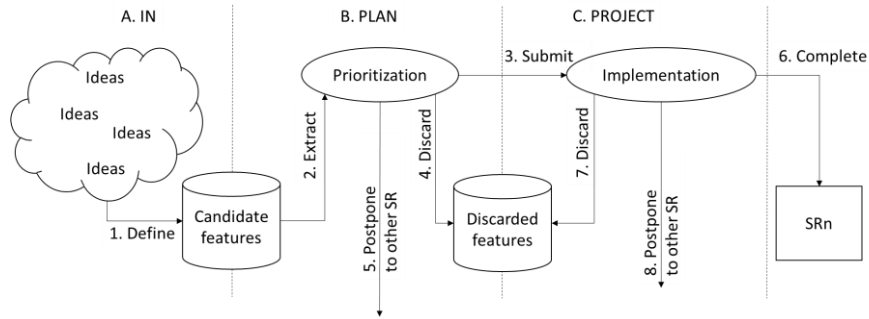

**Fig. 7.** The scope decision process stages for features.

The ratio of QFs proposed by stakeholders that make it all the way to the PROJECT stage in the build-up, new market, and consolidation phases was between 50% and 100%. However, for the growth phase, the median ratio was only 38% of the total of 30 QFs consider in the phase. Testing this with a Wilcoxon signed-ranks test results in a statistically significant difference at $p=0.09$. This implies that QFs were more often dismissed and significant resources were dedicated to architecture refactoring and domain engineering activities. This implication is relevant for RQ2 in that the underlying SPL activities might have an impact on the scope decision process for QFs.

Fig. **8** outlines the details of the number of QFs in the different stages. At some point, the stakeholder will hand over the feature to the product management by defining it in the candidate database and proposing a release of the feature. This signifies a measure of the IN stage QFs, see **Fig. 8**. In the PLAN stage, the product management considers the candidate features and associated release. The product manager can either discard a feature or submit a feature to the next stage. The submitted features signify a measure of features in the PLAN stage that were not discarded; same as for the PROJECT stage. Features completed signifies a measure of features included in a release. Hence, the difference between IN and PLAN is the number of features discarded in that stage; similarly, for PLAN and PROJECT.

After inspection, it becomes apparent that the number of QFs remained relatively low in the build-up and growth phases. The increase is substantial in the new market phase, and most QFs were proposed or realized in the consolidation phase. The median ratio between the number of QFs submitted the project stage that ends up being delivered into the products ("completed") was 50% in the growth phase, 93% in the new market phase and 85% in the consolidation phase. This is in line with opinions from respondents I1 and I4 who perceived the reluctance to accept QFs in the growth phase, that diminished in the new market phase, mainly due to a clear directive from management to include and prioritize QFs into the scope.

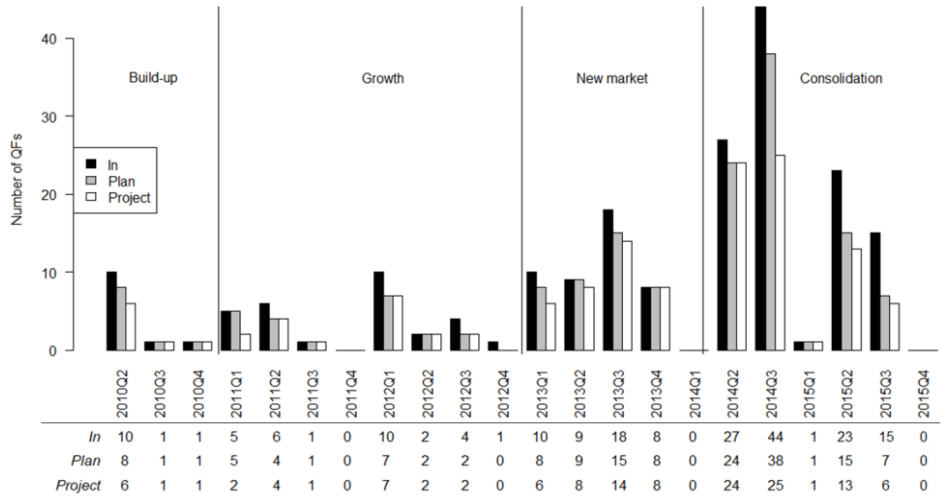

**Fig. 8**. Handling of QFs across the phases and quarters.

We also calculate the ratio between the number of QFs submitted by stakeholders to those that end up in the software product, i.e. are completed. The ratios per phase are 20% for the growth phase, 78% for the new market phase and 48% for the consolidation phase. The median ratio between the number of features submitted by the stakeholders and arriving at the *Plan* stage is 78%, indicating that more filtering is happening later in the process when SPM and projects take on the responsibility of the QFs. This difference in the ratios is statistically significant with the Wilcoxon signed-ranks test results in p=0.1. A possible interpretation of this is that the stakeholders are not good at filtering their own feature or ideas as they might have problems seeing the complete picture, which on the other than the SPM seems to be able to. Another possible explanation is that SPM aggregates various viewpoints from multiple stakeholders and attempts to create the optimal scope from often conflicting demands.

### 5.4   Stakeholders in the scope decision process

**Fig. 9** outlines the QF origin (external or internal stakeholder) during the phases and the quarters. As seen in the figure, the number of QFs proposed by internal stakeholders is small in the build-up phase and the first half of the growth phase. The latter half of the growth phase and in the new market and consolidation phase shows an increase in internally proposed QFs. At the same time, the QFs coming from external stakeholders are about the same. The median ratio of QFs from external stakeholders to the total number of QFs proposed is 100% in the build-up phase, 50% in the growth phase, 33% in the new market phase and 26% in the consolidation phase. When considering the build-up and growth phases together and comparing to the new market and consolidation phases, the difference is statistically significant at p=0.1, using the Wilcoxon signed-ranks test. I4 comments that from the new market phase, customer retention is being heavily discussed and eventually included in the strategic direction of the product line. We interpret the shift from external to internal stakeholders and the increased interest in QFs that the various internal stakeholders more and more consider the customer perspective rather than that of the indirect customers represented by the external stakeholders. I2 corroborates this in that they started using analytics more and more. Hence, it seems as if instead of relying on external stakeholders such as customer accounts, data is elicited directly from the users of the products.

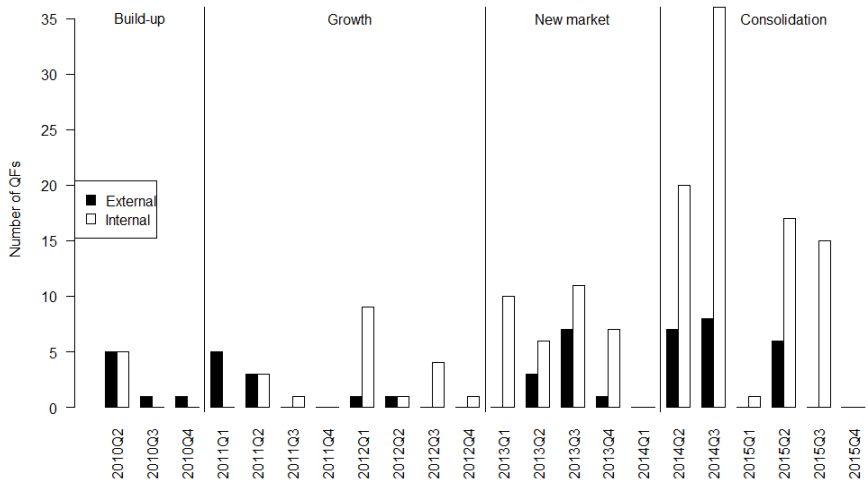

**Fig. 9.** Summary of origin for the QFs per phase and quarter of all proposed QFs.

An analysis of the content of features proposed in releases up to 13Q1 shows an interesting pattern. Completed features were mainly requested by external stakeholders (e.g., customer accounts) who considered them very important and necessary to purchase products from the case company. On the other hand, none of the discarded features (most regarding access to the product, DRM, and IT-security policies) was critical for the external stakeholders or hindering product launch. Looking at security QFs after 13Q1, it appears to be a similar pattern in the decision making of high acceptance ratio of security QFs that are demanded by external stakeholders. This is similar to our previous work where we found that change requests proposed by external stakeholders are more likely to be accepted (Wnuk et al., 2015). Looking at releases 14Q2 to 15Q2, we observe that important external stakeholder QFs are discarded. These QFs are specific for some markets that are not considered important for the case company. Moreover, some QFs that are about to implement solutions available by competitors are also rejected which could indicate that they do not fit into the overall strategy regarding security. Related to RQ3, the data indicates that if the stakeholder role proposing a QF is external, a QF has a higher acceptance rate. However, the data also indicates that there was something missing in the scope decision process in terms of QFs, implying that relying on external stakeholders is insufficient.

## 5.5 Release planning

Interviewee I4 mentioned that at 13Q1, battery performance became part of the strategic scope directive for the product line. The scope directive is created to steer the product line. The scope directive is a document prepared by the strategic planning together with product management at the company. It provides guidelines for scope decision decisions and helps in selecting the features which are in line with the company direction for the upcoming products. The completion ratio of QFs, whether the QA is included in the directive or not, and the FF is presented in **Table 7**. We observe an increasing number of performance features, peaking at the beginning of the consolidation phase. Especially when looking at the implemented QFs, we see an increase from 13Q3, see **Fig. 8**. By 14Q3, the number of performance features started to decrease and by 15Q2 the level is back to the previous. From release 15Q2, performance is no longer part of the strategic direction for the products, which is one possible explanation that the number of QFs is decreasing in **Fig. 8**.

Interviewee I4 observes that internal stakeholders proposed many more *performance* features for battery and those were also accepted throughout the scope decision process. Interviewees I1 and I2 complemented this finding with statements that as the products and markets mature, it is harder to differentiate the offering. Consequently, the importance of QAs, e.g., performance increased for the success of the products. One possible explanation for this significant increase is that the focus before 13Q1 was on customer account input and not direct feedback from usage or consumers. We speculate that, in relation to RQ3, once analytics is collected providing direct feedback from customers, the actual gap regarding performance is evident and there was strategic planning taken place to address the gap.

For the integrity related security QFs, there is a surge starting at 12Q4 and that continues to be strong for new market and consolidation phases, see **Fig. 10**. Interviewee I4 noted that around 13Q2, there was an explicit strategic direction to target a specific market segment. Interviewees I1 and I2 reflected that they got a clear direction to work explicitly towards this market segment. Similar to performance QFs, not all integrity related features had a clear recipient in the development organization. Interviewee I3 explained that this prolonged the introduction of over several releases as the analysis and negotiation was more complex. This pattern is like the performance QFs. We see that for performance, there is a peek around 14Q2 followed by a decline in proposed and accepted QFs. Hence, in relation to RQ2, it appears as if clear strategic input affects the scope decision process and in relation to RQ3 that the same strategic input changes the behavior of the internal stakeholders but not necessarily the external ones.

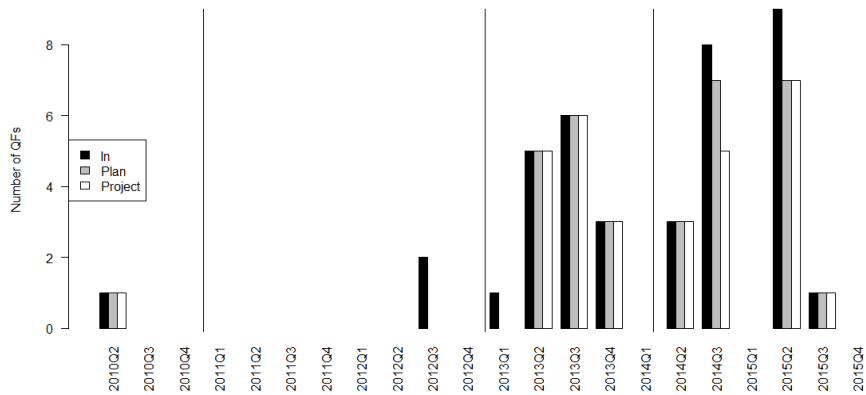

**Fig. 10.** Handling of Integrity QFs across the releases. The bars show the number of features in each stage (cf. **Fig. 7**).

Hence, we observe an impact of the strategic directive across many products, projects, and SPM. From the impact perspective, the battery performance QFs are like the integrity QFs in that they affect several parts of the product line and there is no clear recipient. However, integrity was a part of the scope directive.

**Table 7**
The end state of features and Acceptance ratio across the phases of features submitted to the Plan stage (cf. **Fig. 7**).

|  | In directive | State | Build-up | Growth | New market | Consol-idation | Total |
|---|---|---|---|---|---|---|---|
| QF | No | Completed | 8 | 7 | 28 | 26 | 69 |
|  |  | Discarded | 5 | 11 | 8 | 22 | 46 |
|  |  | Acceptance ratio | 62% | 39% | 78% | 54% | 60% |
|  | Yes | Completed |  |  | 7 | 22 | 29 |
|  |  | Discarded |  |  | 1 | 10 | 11 |
|  |  | Acceptance ratio |  |  | 88% | 69% | 73% |
| All (FF+QF) | N/A | Completed | 212 | 497 | 378 | 929 | 2016 |
|  |  | Discarded | 177 | 404 | 151 | 355 | 1087 |
|  |  | Acceptance ratio | 54% | 55% | 71% | 72% | 65% |

## 5.6 Feedback on the quality levels

As highlighted by Interviewees I1 and I3, there was poor performance in the software for certain QAs, and the attention increased to address them. Interviewee I4 mentions that towards the end of the growth phase, there was a real urgency to address the battery performance issues and around towards the end of the new market phase, user experience issues QAs got to the top of the strategy. Even though the development projects continuously tried to improve battery performance, this was insufficient without a clear strategic direction. Interviewee I3 highlighted that battery issues are truly cross-cutting the entire product line. As there was no clear development group or team to analyze and solve these issues, they become much more difficult to handle from an organizational perspective. All interviewees agreed that battery performance is the single most important quality aspect of the products. However, they also stated that the strategic process failed to explicitly address poor battery performance.

The company struggled to appropriately handle the performance issues, as noted in the interviews. One of the problems was a lack of feedback in the scope decision process. Despite the customers signaling insufficient quality levels, the strategic and operational decision processes were reacting too slow. Interviewees I2 and I4 commented that the battery times were getting acceptable, which meant that even better battery performance would not give a competitive advantage. Hence, the deficiencies built up in the build-up and the growth phases seem to be addressed during new market phase to a large extent.

During the consolidation phase, we see a bigger rejection of performance related QFs. Internal stakeholders still propose many new performance QFs, but SPM does not accept them with the same frequency. This indicates that there is a lack of alignment between the stakeholders and the SPM. On the other hand, the SPM and development teams seem aligned, as the development teams still accept most of the features requested by SPM. That is, QFs are primarily discarded in the PLAN stage by the SPM and not in the PROJECT stage. Our interpretation is that the operational work by the SPM and development project are well aligned but the strategic alignment across organizations is deficient.

Interviewees I2 and I3 noted that there was an explicit strategic direction to improve the user experience during the consolidation phase. The background is consumer data indicating large dissatisfaction with the user experience in several areas. I.e. customer services data and market research triggered the direction of the strategy. In total over the studied period, 26 QFs related to usability were proposed, making usability the third largest category of QFs constituting 13% of the total number of QFs. Half of the usability QFs were proposed after 14Q2. Hence, in relation to RQ2, both the strategic direction and consumer data impacts the scope decision process.

As opposed to the other QFs, no external stakeholder was requesting usability QFs, since only 2 of the 26 usability features originated from external stakeholders. Many of the rejected features are rejected not because they lack value. Rather, the decisions on usability and how to address them are handled on a more detailed level by feature teams instead of in the scope decision process on a feature level. This implies, in relation to RQ3, that the stakeholder role – internal or external – is related to which QAs they propose.

## 6 Discussion and future work

### 6.1 RQ1: Are quality aspects explicitly specified for features??

Research question **RQ1** aims to study whether quality aspects are handled in the scope decision process. Of the 4446 features included in the study, 196 (4.41%) are specifically addressing a quality aspect. In previous work, we have seen that as many as 38% of the requirements in a vendor specification were QRs (Berntsson Svensson et al., 2013). One explanation for the large difference between the two studies could be that the abstraction level is different (feature level vs. requirements level). Features usually appear earlier in the process than requirements (requirements are usually a result of detailing, analyzing or refining features), one possible explanation could be that quality aspects remain hidden within functional features and that customer expresses mostly needs and goals associated with functionality, leaving quality aspects and levels to be later decided. Furthermore, it might also be the case that the QRs never show up as formal requirements, as a survey by Ameller et al. suggests (Ameller et al., 2013). They are instead handled by the software developers. Several researchers stressed negative consequences of this behavior (Berntsson Svensson et al., 2012; Yu, 1997) and advocated that reasoning about QRs should be introduced earlier in the process and supported by goal modeling (Dardenne, van Lamsweerde, & Fickas, 1993). Another possible explanation is since the company currently does not perform any form of modeling or reasoning for quality in early requirements engineering phases (Yu, 1997), quality is left for later phases and does not manifest itself among the features.

Similar to Berntsson Svensson et al., (Berntsson Svensson et al., 2012), the data suggest that FF are more important than QFs. The acceptance ratio is slightly lower and the quantity of QFs is much smaller (cf. **Table 7**). Quantity does not equate to importance, however. Several of the interviewees did, however, point out deficiencies in the QFs handling. Hence, we speculate that the handling of QFs warrant improvement. A difference between our study and Berntsson Svensson et al. is the handling of performance QF; they report that they are more often dismissed for B2C which our data does not suggest. One possible explanation is that the survey answers from Berntsson Svensson et al. might report a wanted state, which is not necessarily reflected in the actual operations. There might also be a different interpretation of performance from an ISO 25010 perspective. We interpret this in two different ways: 1) There is still a lack of understanding of the actual needs and practices in industry 2) There is a need to understand QFs and QRs in a more general strategic level. We speculate that there might be a difference in the competence in handling QFs, similar to what is reported from the automotive domain (Weber & Weisbrod, 2002) and that industry still needs support for early identification of quality goals and aspects.

We believe there is a need for explicit scope decisions on both strategic and operational levels. Deciding on the software scope for several parallel releases is a continuous activity (Wnuk et al., 2016). We speculate that had there not been a scope directive in place, then the handling of the Integrity QFs would have looked more like that of the battery QFs. We also see that the acceptance ratio for QFs being part of the scope directive is higher than for those QFs not part of it (cf. **Table 7**). Finally, the interviewees point out that without the coordinated efforts because of the scope directive, the performance issues would have taken even longer to handle. Even though the Software Product Management literature recognizes the need for planning across several releases (Kostoff & Schaller, 2001; Regnell & Brinkkemper, 2005; Rinne, 2004; Vähäniitty et al., 2002) and setting quality levels for candidate features (Regnell, Svensson, & Olsson, 2008), there is a lack of operational support for incremental and iterative planning of QRs across several releases.

## 6.2 RQ2: What does the scope decision and scope decision process for quality features (QFs) look like across the product line lifecycle?

That the number of QFs vary over time is hardly surprising. For example, in the build-up phase, it is not surprising that the focus is on getting to the market and maximizing the number of FFs. The data and the interviewees indicate that QFs get less focus in the build-up and growth phases. This appears to be logical due to a strong focus on quickly delivering products to the market strengthens the dominance of FRs over QRs in the phase where the expansion of the number of products and market shares is the strongest. This finding is similar to what has been observed in agile RE, where a challenge is to properly handle QRs (Inayat, Salim, Marczak, Daneva, & Shamshirband, 2015). Ernst and Mylopoulos report a study of two OSS projects, where the main result point to a decreasing interest in QF rather than increasing (Ernst & Mylopoulos, 2010). This seems contradictory to our findings, as we see fewer QFs in the build-up. In their study, they use communication logs and comments in the tool for their analysis whereas we use more formal documents and interviews.

The acceptance ratio of QFs in the growth phase is lower – almost half – than in the other phases (cf. **Table 7**). The number of QFs both in the build-up and growth phase is however low; small changes largely influence the result. The interviewees observe that in the growth phase, there was a focus on preparing the SPL for a larger product portfolio where the QFs were not considered key. At the same time, during the relatively long growth phase, not many QFs are proposed. Hence, the entire organization is not focused on QFs. Lastly, at the time, the market was expanding rapidly and technology changing fast. Hence, there might have been less interest from the market in QFs. We speculate that there was a lack of foresight into the importance of QFs and that measures could have been taken to have a better preparedness for QFs in the coming phases.

A practical implication here is that the failure to incorporate QFs early in the build-up phase may cause delays in addressing them until after the growth phase since the acceptance ration in the growth phase is low and the focus is on architecture refactoring. Companies should be aware of that and intensify QFs strategic planning as early as possible to ensure to ensure that the relevant QFs are included in the scope planning (Borg et al., 2003; Ebert, 1998; H.-B. Kittlaus & Clough, 2010). At the same time, delivering sufficient quality levels in the product is important for product success (Ebert, 1998; H.-B. Kittlaus & Clough, 2010). QFs should be managed from the start of a product line to ensure that proper decisions related to product success are made (H.-B. Kittlaus & Clough, 2010). Furthermore, companies should strive to capture and consider these dependencies when estimating effort and release planning of QFs (Ruhe & Saliu, 2005). Finding the right balance between quality and functionality while not exceeded software development capabilities is an important area for further investigation.

It seems that the SPM is well aligned both with (internal) stakeholders proposing QFs and the projects implementing them (cf. section 5.6). However, even though they could align with the existing organization and division of subject areas, it is more problematic when addressing cross-cutting concerns. Furthermore, QFs require several releases to be addressed. Depending on the products, the markets, consumers, etc., the priority between FFs and QFs and among the different QFs differ. Hence, it is too naïve to assume that e.g., performance is always down-prioritized by B2C. Furthermore, both technological as well as organizational, it can take time to shift the focus from FFs to QFs or to different QFs. Hence, deliberate actions are required in order not to be reactive and slow in response to customer feedback.

We hypothesize that there is a need for an explicit feedback from product usage data to the scope decision process, in combination with internal stakeholders (such as experts) and customer accounts input. Some QFs, such as performance, are difficult to estimate upfront and they are perceived differently by different users. Hence, for performance, it seems to make sense to work in a feedback manner. However, for security, our results suggest the opposite. Even though some aspects of e.g., integrity can be analyzed from actual usage, many parts can just as easily be planned up front.

## 6.3 RQ3: How do different roles influence the decision for quality features (QFs) over the scope decision process lifecycle?

We investigated the role in the scope decision process and how they were changing over time. We conclude that external stakeholders are more prominent as a source for QF at the beginning of the lifecycle. As the work progresses, two things happened: 1) more and more the internal stakeholders are proposing QFs towards the end of the growth phase 2) the externally proposed QFs are mostly rejected in the new market phase. What is also clear is that product usage data and direct consumer feedback is not explicitly utilized in the scope decision process. The company is collecting both general business intelligence such as market information, competitive intelligence, etc., and later in the product line lifecycle also product usage data ("analytics"). However, the interpretation of these is left to experts and not directly channeled into the scope decision process. Furthermore, there are indications that the need for a quicker feedback process. At the same time, this needs to be balanced with the normal plan-driven process, as seen for example with the integrity related QFs.

In terms of types of QFs, performance is the most prevalent followed by security and usability (see **Fig. 4**). Berntsson Svensson et al. found in the B2B domain, where safety, performance, and reliability are the most important QAs. Furthermore, they report that for the B2C domain, the most important QAs are usability, performance, and stability. A report from the medical domain indicates that usability and interoperability are the most important QAs, followed by security and performance (Defranco, Kassab, Laplante, & Laplante, 2017).

A too strong reliance on external stakeholders for QFs did not bring a balanced picture of the state of quality for the actual end-consumer. Both elicitation (i.e. creating new features) and prioritization (i.e. scope decision) turned out to provide limited efficiency. This implies that getting as close as possible to the actual end-consumers is important to understand the actual QFs needed.

We hypothesize that the prevalence of performance and security features in our case study is related to the strong customer account representatives, driving a B2B perspective even though the products are B2C oriented. At the same time, speculate that the lack of usability QFs is an issue for B2C. Therefore, this indicates that there is a need to balance the input from different types of stakeholders and highlights a need to have a strategy for which data is in the scope decision decisions. More efforts should also be directed towards exploring the relationships between the stakeholder roles and their involvement in decision making (Wnuk, 2017).

Business intelligence data and product usage data should explicitly be connected to the scope decision process rather than being processed by stakeholders. If the information is passed along too many times, it will take longer before actions can be taken and the original meaning might be lost along the way (Bjarnason, Wnuk, & Regnell, 2011). If the data scope decisions are too disconnected from the product scope decisions, there is a risk of misalignments.

## 6.4 Validity discussion

We discuss threats to validity based on the classification suggested by Runeson et al., (Runeson et al., 2012).

*Construct validity* concerns with the relationship between the theory and the observation. Two operational measures (archival analysis of scope decision history and interviews) were selected to minimize the risk of missing important perspective on the studied phenomenon. Moreover, two researchers attended every interview session to reduce misinterpretations by the researchers and the interviewed persons. The exploratory nature of our study limited our expectations of the obtained results as no theory was suggested before the study. We have also minimized reactive bias as two researchers have a long research collaboration with the case company and explained confidentiality rules and avoided assumed expectations. On the other hand, one of the researchers also has considerable experience with the company in question, and hence potentially a bias to some conclusion. We address this by having a very clear process and reporting all process steps. We minimized mono-operation and mono-method bias by combining archival analysis with interviews, validating the interview instrument and independently categorizing the features. We selected interviewees based on their extensive experience in working with the scope decision process and therefore their responses have high creditability. Finally, multiple researchers validated the interview instrument to avoid misinterpretations.

*Reliability* deals with to what extent the data analysis and collection are dependent on a specific researcher and the ability to replicate the study. We enhanced reliability by documenting each step of the scope decision database analysis. To ensure a consistent categorization, two researchers were involved in the process. By iterating the coding several times until we agreed on categorization. There is still a risk that, even though we agree on the categorization, that we might both be wrong. However, based on a good initial interrater agreement, we believe that despite the risk of our categorization converging to incorrect categories is low in the iterative process.

Lead-time data for 650 features (both FFs and QFs) were excluded from the analysis as we judged the data to be incorrect. The main indication for incorrect data was a very long lead-time for a feature. This threats validity for the lead-time analysis since we should have 1) removed features from the lead-time analysis even though they should have remained (false negative) and 2) kept features with incorrect lead-time, though shorter than 365 days (false positive). 650 of 4446 features were excluded. Hence, false negatives should have a minor impact. Furthermore, keeping true negatives would have significantly skewed the results, making the lead-time analysis meaningless. The excluded features found across all the releases, further reducing the risk of impacting the result. It is more difficult to estimate the risk of false positives. For the analysis of lead-times across phases, the threats to reliability remain low. However, there is a risk that the difference between discard and completed might change if we would have included the 650 features. We are addressing this by being more careful in conclusions related to the difference between completed and discarded feature lead-time.

We transcribed the interviews and described our interviewees to enable further analysis or replications. We allowed our interviewees to review and comment on the results of the quantitative analysis. There is a threat that our interviewees do not have a complete understanding of the scope decision process. However, as the interviewees have considerable experience with the scope decision process from complementing perspectives, we believe the risk of incorrect results is low. Furthermore, the interviewees and data are, on a general level, aligned, indicating that our interpretation is correct. In terms of statistical tests, we used non-parametric tests that do not assume a normal distribution. The anonymized results are documented together with the analysis steps to enable replications and can be obtained upon request.

Finally, we used multiple sources of data (artifact analysis and interviews) and performed observer triangulation during the interviews. The former reduces the risk of skewed data coming from one source. The latter mitigates researcher bias in interpreting the interviews.

*Internal validity* concerns with possible confounding factors that may impact the studied causal relationships. This case study is exploratory rather than explanatory or confirmatory and thus confounding factors are minimized. Moreover, we have not tested any causal relationships and threats to internal validity can be considered as having minimal impact in this case. Whenever cause-effect relationships are discussed in the results and analysis section, they should be interpreted with caution. Moreover, although we introduced some subjectivity in our dataset by asking practitioners to provide missing information for a small number of features, we consider this threat to have minimal effect on the results.

*External validity* concerns whether the results are applicable outside the case in question. We used analytical generalization rather than statistical and presented the case study details (Flyvbjerg, 2006; Yin, 1994). The selected case company is a large organization with an intense focus on software development for embedded B2C devices, regularly releasing software updates. We believe that it is reasonable to assume that the results can be useful to companies with a similar context and setup. Moreover, we strengthen internal generalizability by analyzing extracting over 7000 features and analyzing over 4000 features available in the feature database at the time of the study and selecting different roles as interviewees. The fact that we analyzed 36 software releases that involved over 60 products between 2010 and 2015 brings additional confidence in the representability of our dataset and strengthens external validity. However, generalizing the findings to any other company is not possible. However, the difficulty to generalize should not be exaggerated (Flyvbjerg, 2006). We believe that, through our case description, similar companies should be able to learn from our case study and similar phenomena as we have found are likely to be present.

## 6.5 Future work

We are currently working towards a conceptual model for scope decisions on QFs (Olsson & Wnuk, 2018). We plan to validate both the underlying empirical findings of this study as well as to validate the applicability of our conceptual model with several companies. plan to compare the QFs scope decision process and whether the results of this study are generalizable to other cases. We also invite other researchers to replicate our results to even further improve generalizability.

Another track we are considering is applying the approach from Ernst and Mylopoulos with other sources than open source software projects (Ernst & Mylopoulos, 2010). Specifically, if a technique could be developed to dynamically mine customer service and social media data for quick feedback to the scope decision process, we believe this has a potential to complement the current trends of experimentation. On feedback loop: With a mindset similar to "You are a support person" in Parnin et al., (Parnin et al., 2017), working with product usage data and enabling direct product usage feedback into the decision-making processes along with other data from actual users could have a potential to speed up the scope decision process.

Interestingly, in our data, there is no significant difference in the lead-time for quality and functional features. This could imply that there is no need to have a separate scope decision process for QFs. However, interestingly, the lead-time for features which end up being discarded is larger than that of those who end up in the software. Looking at the data, it seems that there are two main contributors to the lead-time of discarded features: SPM decision and integration decision. The latter implies that there is in fact implementation work performed, even on discarded features. Unfortunately, we do not have data at this level of granularity for detailed statistical analysis. It would be interesting, however, to study this in more detail and see if it is possible to identify predictors of features which will be rejected at a later stage, to avoid spending effort in them and to avoid creating expectations on the scope.

There are not many studies on scope decision of QFs over a long period. We believe there is a need for further work to understand the key factors influencing the scope decision process. Even though we believe the results of this study is interesting and relevant, we would like to study other organizations to test whether our findings are generalizable to other organizations. We plan to perform further case studies on other companies and are also planning a survey to reach a wider range of companies and respondents. Finally, we also see a need to study interdependencies among QRs and the impact on scope decisions both on an operational and a strategic level.

# 7   Conclusions

The product success depends on what functions the product performs but also how well the product performs it. The subjective, relative and interacting nature of QRs makes it challenging to reach an agreement among the relevant stakeholders on what to and when. This paper focuses on how scope decision decisions are made for QRs on a feature level considering data from a period of five years, spanning several products and releases of the software.

We observe that the number of features explicitly addressing quality is quite low at the beginning of the product line lifecycle and that increases after about three years. However, the number of QFs is still small in comparison to FFs. From the interviews and the data analysis, we observe the clear need to address quality sooner and more explicitly, but the company was not set up to promptly react. This is especially evident in battery performance. Once the case company was able to change the organization to focus on battery performance, they were able to systematically address the quality issues. The challenges seem to lay in being able to establish long-term strategic planning and at the same time increase responsiveness to customer feedback and market changes.

Our results also indicate that a too strong reliance on upfront analysis and external stakeholders as a source for features lead to a slow reaction to key market trends and feedback. However, scalability becomes an issue when developing products for a global market with thousands of involved software engineers, (Regnell, Svensson, & Wnuk, 2008). Hence, there is also a need for a top-down, forward feeding strategic process to guide and steer all the teams in an aligned direction. As reported by Bjarnason et al., the lack of vision and clear goals is a hindrance to the communication and hence contributes to unfulfilled expectations on quality (Bjarnason et al., 2011). The key hypothesis we propose is to ensure alignment of long lead-time, forward driven processes with short cycles of feedback-driven and agile processes. We are currently developing and evaluating a model to support a flexible and adaptable scope decision process combining a strategic and long-term perspective with an agile and short-term decision-making approach.

**Acknowledgment.** We want to thank all the participants in the interviews. This work is supported by the IKNOWDM project (20150033) from the Knowledge Foundation in Sweden.

# Appendix

## A. Interview guide

The purpose of this interview is to look into how Quality requirements (QR) are handled on a scope level. Features from the platform have been analyzed and categorized, i.e. a document analysis. The purpose of the interview is to give more background and qualitative information to explain and motivate the findings in the document analysis. [2]

### Section 1

Introduction and background data on the interviewee. Keep in mind that we are taking about a timeframe from 2010-2015.
1. How long have you been working in the company and what have been your roles?
2. What has been your roles during the studied timeframe?
3. What is your experience working with QR? [Keywords to look for: definition, analysis, verification, specification, role in scope process]
4. Have you worked with or close to scope management? [Keywords to look for: scope decisions, SIA, planning]

### Section 2

Dig into QR and product strategy. Purpose is to first with open questions get the interviewee's perception of how the company has worked with QR over time. Also, try to fill in the blanks in terms of missing strategy input.
5. Considering since you started with the platform, how has the meaning and importance of QR changed over the course of time? Which QR have been important?
    a. Start open discussion to find out
        i. If and why QR are important and how that has changed over time?
        ii. What is the driver for QR? Rationale? Why?
        iii. Which type (ISO) of QR are important? How has this changed?
        iv. How has priority of QR in the scope process changed since you started with the platform? Why?
        v. Which stakeholders typically drive QR? Why?
        vi. Especially, has there been strategic input (formal or informal)? How?

---

[2] The text has been edited for the sake of presentation and anonymity.

vii. Has the process/governance impacted?
                b. Look at some of the charts and start discussing the main releases. Try to cover same topics as 5a.
                        i. Look at the overview of QFs in the scope decision process. What happened? Why?

What happens around SR22 and SR30?
Before SR22, QR not important?
Look at the types of ISO (50% performance, 25% security)
                        ii. Look at accept/reject
What happens after the peak SR30?
                        iii. Talk about directive [show scope directive] What other strategic input is there? What existed before the directive? Any type of strategic important of QR before directive? Especially
                                1. SR25-SR30 (priority, fill in strategy)
                                2. SR14-SR21 (priority, fill in strategy)

## Section 3

Open discussion
  6. What has worked well with QR? Problems?
  7. What is happening now with QR?
        o Are they prioritized for future releases?
        o How do you work with them?
        o Strategic input?